\documentclass{preprint}

\usepackage{lipsum} 
\usepackage{xcolor}
\usepackage{booktabs}
\usepackage{graphicx} 
\usepackage{xspace}
\usepackage{ulem}
\usepackage[colorlinks=true]{hyperref}
\usepackage{threeparttable}
\usepackage{tabularx}
\usepackage{textcomp}

\title{Improving Fairness of Automated Chest X-ray Diagnosis by Contrastive Learning}
\author[1]{Mingquan Lin}
\author[2]{Tianhao Li}
\author[1]{Zhaoyi Sun}
\author[2]{Ying Ding}
\author [3] {Gregory Holste}
\author[1] {Fei Wang}
\author[4]{George Shih}
\author[1,*]{Yifan Peng}
\affil[1]{Department of Population Health Sciences, Weill Cornell Medicine, New York, USA}
\affil[2]{School of Information, The University of Texas at Austin, Austin, TX, USA}
\affil [3] {Department of Electrical and Computer Engineering, The University of Texas at Austin, Austin, TX, USA}
\affil[4]{Department of Radiology, Weill Cornell Medicine, New York, USA}
\affil[*]{Corresponding: yip4002@med.cornell.edu}

\newcommand{\dmauc}{$\Delta$mAUC\xspace}
\newcommand{\dtpr}{$\Delta$TPR\xspace}
\newcommand{\dfpr}{$\Delta$FPR\xspace}
\newcommand{\dbs}{$\Delta$BS\xspace}

\begin{document}

\maketitle

\begin{abstract}
\textbf{Purpose}:  Limited studies exploring concrete methods or approaches to tackle and enhance model fairness in the radiology domain. Our proposed AI model utilizes supervised contrastive learning to minimize bias in CXR diagnosis.

\textbf{Materials and Methods}: In this retrospective study, we evaluated our proposed method on two datasets: the Medical Imaging and Data Resource Center (MIDRC) dataset with 77,887 CXR images from 27,796 patients collected as of April 20, 2023 for COVID-19 diagnosis, and the NIH Chest X-ray (NIH-CXR) dataset with 112,120 CXR images from 30,805 patients collected between 1992 and 2015. In the NIH-CXR dataset, thoracic abnormalities include atelectasis, cardiomegaly, effusion, infiltration, mass, nodule, pneumonia, pneumothorax, consolidation, edema, emphysema, fibrosis, pleural thickening, or hernia. Our proposed method utilizes supervised contrastive learning with carefully selected positive and negative samples to generate fair image embeddings, which are fine-tuned for subsequent tasks to reduce bias in chest X-ray (CXR) diagnosis. We evaluated the methods using the marginal AUC difference (\dmauc).

\textbf{Results}: The proposed model showed a significant decrease in bias across all subgroups when compared to the baseline models, as evidenced by a paired T-test (p$<$0.0001). The \dmauc obtained by our method were 0.0116 (95\% CI, 0.0110-0.0123), 0.2102 (95\% CI, 0.2087-0.2118), and 0.1000 (95\% CI, 0.0988-0.1011) for sex, race, and age on MIDRC, and 0.0090 (95\% CI, 0.0082-0.0097) for sex and 0.0512 (95\% CI, 0.0512-0.0532) for age on NIH-CXR, respectively.

\textbf{Conclusion}: Employing supervised contrastive learning can mitigate bias in CXR diagnosis, addressing concerns of fairness and reliability in deep learning-based diagnostic methods.

\end{abstract}

\section{Introduction}

In recent years, Artificial Intelligence (AI) has been extensively utilized in image-based disease diagnosis.\cite{Yu2018-wz, Lin2021-sr, Wang2017-ChestX, Rajpurkar2017-se, Lin2022-uw, Lin2022-jo} While these models have attained or exceeded expert-level performance, the concern of fairness has emerged in various medical domains and populations.\cite{Char2018-fu} In the AI algorithm, fairness denotes the absence of bias or favoritism towards an individual or group based on their inherent or acquired characteristics.\cite{Mehrabi2021-dp} In medical domains, certain groups, such as those defined by race, sex, and age, have been identified as being subject to unfair or biased decisions made by AI models.\cite{Lin2023-hv, Seyyed-Kalantari2021-dn, Lin2023-wq}

A chest X-ray (CXR) is a quick and convenient diagnostic tool that uses a low dose of ionizing radiation to produce images of the chest, including the lungs, heart, and chest wall. This imaging technique can shed light on the underlying cause of shortness of breath, persistent cough, chest pain, and injury. Additionally, CXRs help diagnose and monitor lung conditions such as pneumonia, emphysema, and cancer. Several studies have focused on automating disease diagnosis based on CXR imaging to achieve accurate results.\cite{Abbas2021-yk, Minaee2020-xr, Liu2023-vq, liu2023new} While these efforts have achieved high accuracy in detecting abnormalities in CXRs, exploring AI model fairness and bias reduction has been relatively limited. Therefore, there is a need to develop methods to minimize bias in CXR diagnosis.

There exist three primary methods to reduce bias in medical image classification. Pre-processing methods work to reduce bias through dataset resampling or augmentation.\cite{Seyyed-Kalantari2021-dn, Joshi2021-ng} In-processing methods typically incorporate an adversarial component into the baseline model. This component predicts sensitive attributes derived from the input image and emphasizes the loss function selection.\cite{Zhao2020-zf, Du2023-eb} Lastly, post-processing techniques can address unfairness by introducing perturbations to input images.\cite{Wu2022-ga} This prevents the model from relying on biased features and can be achieved without necessitating model retraining. Despite these strategies, they have two limitations. First, the changes might inadvertently affect overall performance. This type of degradation, where fairness is achieved by deteriorating the performance of one or more groups is quite problematic.\cite{Zhang2022-rr, Pfohl2021-ub} Furthermore, the testing and development of these methodologies are predominantly conducted on relatively small datasets. This can impede their ability to be generalized or applied to more extensive, real-world scenarios.

Our study aims to investigate fairness issues in employing AI for CXR diagnosis and to mitigate biases on subgroups such as race, sex, and age. One potential reason for bias in AI models is the presence of non-neglectable subgroup information in image embeddings. For example, in the race subgroup, the image embeddings may contain race-related information that could lead to biased predictions by the models. Supervised contrastive learning is a pretraining technique that uses label information to draw embeddings from the same class closer and push those from different classes further apart.\cite{Khosla2020-kq} Benefiting from well-trained embedding, it achieves superior performance on downstream classification tasks. Inspired by this method, we propose utilizing supervised contrastive learning with carefully selected positive and negative samples to generate fair image embedding. Subsequently, the model is fine-tuned for downstream tasks. In our approach, we define images with the same label from different subgroups as positive samples and images with different labels from the same subgroup as negative samples. This demonstrates a significant capability to reduce bias for subgroups.

\section{Materials and Methods}

The study protocol was approved by the institutional review board at each clinical center and Weill Cornell Medicine. Due to the publicly available nature of both datasets used in this study, the requirement for obtaining written informed consent from all subjects (patients) was waived by the IRB.

\subsection{Dataset Acquisition}

\begin{figure}
    \centering
    \includegraphics[width=.5\linewidth]{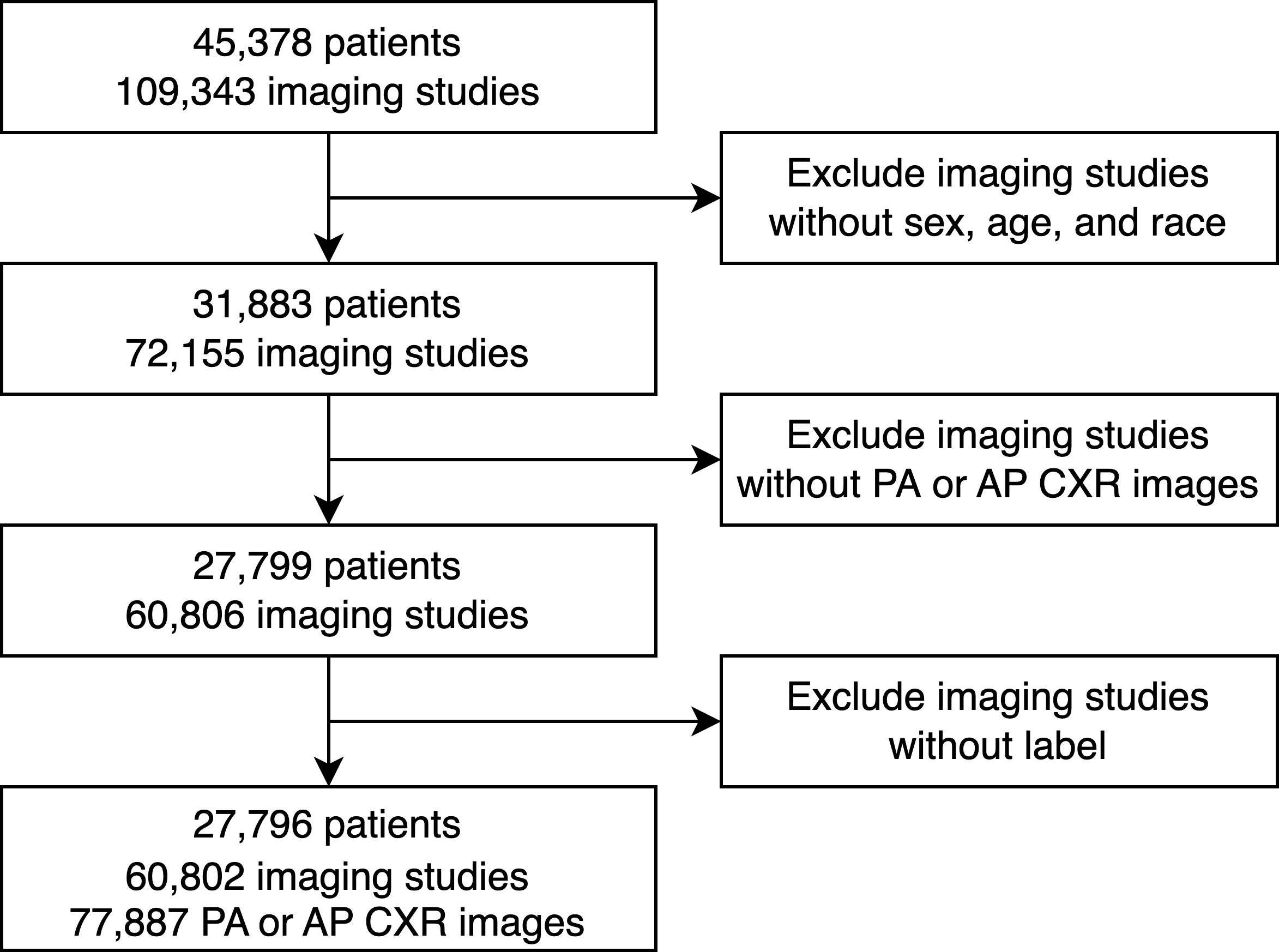}
    \caption{Creation of MIDRC dataset}
    \label{fig:midrc}
\end{figure}

Our proposed method was designed and assessed using two CXR imaging datasets. The first dataset is a repository created for COVID-19 diagnosis, hosted at the University of Chicago as part of the Medical Imaging and Data Resource Center (MIDRC).\cite{Lakhani2023-lp} The MIDRC is a collaborative initiative funded by the National Institute of Biomedical Imaging and Bioengineering (NIBIB) under contracts 75N92020C00008 and 75N92020C00021 and jointly led by the American College of Radiology\textregistered~(ACR\textregistered), the Radiological Society of North America (RSNA), and the American Association of Physicists in Medicine (AAPM). The MIDRC accepts images in DICOM standard and clinical data in various formats. It is currently seeking COVID-19-related CT scans, X-rays, MRI, and Ultrasound, along with similar control cases. In this study, we focus on X-rays. The race, sex, and age data in MIDRC are self-reported. According to the MIDRC Data Contributor Reference Document, the outcome in MIDRC was confirmed through COVID-19 test results (PCR or Rapid antigen test) within a timeframe of 0 to 14 days before the imaging study. Given that MIDRC is a multi-institutional collaborative initiative, and no exclusion criteria are specified in the dataset descriptions, we have taken measures to mitigate selection bias. However, it is important to acknowledge that the MIDRC may not fully represent all patient populations. As of September 2022, there are 126,295 imaging studies with demographic information in the MIDRC data. We collected computed radiography (CR) and digital radiography (DX) with age, sex, and race information. Figure \ref{fig:midrc} provides an overview of the data selection process. Finally, there are 77,887 CXR images from 60,802 imaging studies of 27,796 patients.

The second data set used in this study was the publicly accessible National Institutes of Health Chest X-ray (NIH-CXR) dataset, which comprised 112,120 frontal CXR images from 30,805 patients.\cite{Wang2017-ChestX} In the NIH-CXR dataset, race and sex information were self-reported, while age was recorded at the time of the patient's first admission. In the NIH-CXR dataset, a thoracic abnormality refers to any abnormal finding in the chest area. This encompasses various conditions, such as lung masses, among others.

\subsection{Bias Definition}

To assess the model's fairness, we used the \textbf{\underline{D}}ifference between the maximum and minimum value of the \textbf{\underline{m}}arginal overall \textbf{\underline{A}}rea \textbf{\underline{U}}nder the receiver operating characteristic \textbf{\underline{C}}urve (\dmauc). The marginal AUC\cite{Narasimhan2020-fp} is defined as

\begin{equation}
A_{G_{i>}} := P\left( f(x) > f(x^{'}) \middle| y > y^{'},(x,y) \in G_{i}^{+},(x^{'},y^{'}) \in G^{-} \right)
\end{equation}

\(G\) is the dataset used, \(G_{i}\) is the subgroup in the dataset, \(f(x)\) is the output of the AI model with input image \(x\), and \(y\) is the ground truth label for \(x\), indicating whether the input image has a disease or not. \(P\) stands for the marginal AUC, which measures the AUC for a specific subgroup. It is calculated by determining the probability that the model ranks a randomly selected positive sample from the subgroup, over a randomly selected negative example from the entire data. For binary classification, the marginal AUC requires that positive labels have an equal chance to be predicted positively across subgroups.\cite{Narasimhan2020-fp} By subtracting the minimum value of marginal AUC from the maximum, \dmauc can be obtained. A higher \dmauc signifies significant disparities at the levels of individual subgroups and a lack of fairness in the model's predictions. For example, in the age subgroup, the marginal AUC for individuals below 75 years and their counterparts is 0.8288 and 0.7289, respectively, resulting in a \dmauc of 0.0999. If the proposed method can reduce this value from 0.0999 to a lower value, it successfully reduces bias.

Additionally, we also use the difference between the maximum and minimum values of the subgroups in traditional evaluation metrics, specifically true positive rate (TPR), false negative rate (FPR), and brier scores (BS) to assess fairness. We refer to them as \dtpr, \dfpr, and \dbs.

\subsection{Overall Architecture}

\begin{figure}
    \centering
    \includegraphics[width=.6\linewidth]{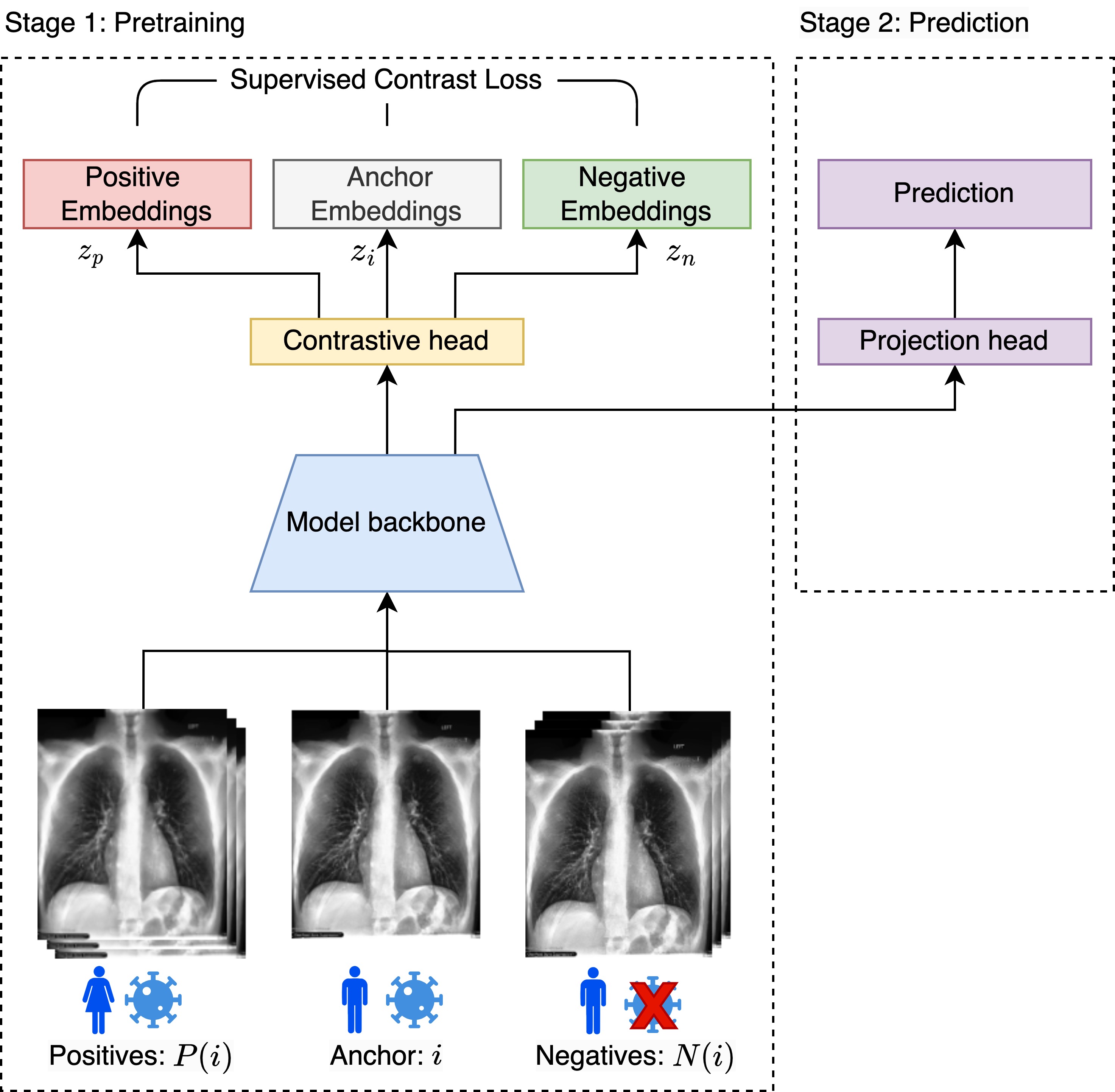}
    \caption{The overview of the proposed workflow using the Contrastive Learning Model for Fairness. For example, a male with COVID-19 serves as the anchor image, while the image of a female with COVID-19 that follows also serves as a positive sample. On the other hand, the image of a male without COVID-19 is considered as a negative sample.}
    \label{fig:overview}
\end{figure}

Our overall architecture is presented in Figure \ref{fig:overview}. We first pre-trained the model using contrastive learning, which learns the initial parameters for the model backbone. We then fine-tuned the model for the subsequential tasks. We used the DenseNet-121 as the backbone in this study.\cite{Huang2017-xq}

\subsection{Contrastive Learning Model}

We use contrastive learning as a pretraining technique to minimize the bias among different subgroups, resulting in fair image embeddings. To implement contrastive learning, we replace the final output layer of the prediction network with a single-layer perceptron, which serves as the contrastive head. In contrastive learning, an "anchor image" refers to an image that serves as a reference point within the contrastive loss function. For anchor images in a minibatch, we use images that have the same label but originate from different subgroups as positive samples, and images with different labels but from the same subgroup as negative samples. In this scenario, a male with COVID-19 serves as the anchor image, while the image of a female with COVID-19 that follows also serves as a positive sample. On the other hand, the image of a male without COVID-19 is considered a negative sample. In this context, positive sampling encourages image embeddings from different subgroups to be similar to one another, while still considering the label information. Image embeddings are the feature embeddings obtained by the convolutional part of the model. Conversely, negative sampling pushes image embeddings with distinct labels further apart, without emphasizing the group information. The contrastive loss can be expressed as follows.\cite{Khosla2020-kq}

\begin{equation}
L = \sum_{i \in I}^{}{\frac{-1}{|P(i)|}\sum_{p \in P(i)}\frac{\exp^{z_{i} \cdot z_{p}/\tau}}{\sum_{n \in N(i)}^{}\exp^{z_{i} \cdot z_{n}/\tau}}}
\end{equation}

where \(i\) is the anchor image in the minibatch \(I\). \(P(i)\) represents all the positive samples of \(i\) in the minibatch. \(N(i)\) are all the negative samples of \(i\) in the minibatch. \(z_{i}\), \(z_{p}\), and \(z_{n}\) are the image embeddings of \(i\), \(p\), and \(n\). The loss function allows all positive pairs to contribute to the numerator, encouraging the encoder to provide closely aligned representations for all entries from the same class. The form of the loss function can distinguish between positive and negative samples.

\subsection{Downstream Prediction}

After we pre-trained the model using contrastive learning, we replaced the contrastive head with the origin output layer, which is the prediction head in Figure \ref{fig:overview}. We then fine-tuned the model to generate the output result. We used binary cross-entropy loss in the downstream prediction.

\subsection{Experimental Settings}

For the MIDRC dataset, we followed the same image processing method as described in the study by Johnson et al.\cite{Johnson2019-rn} for the original CXR images. We started by converting all the Posterior-Anterior (PA) or Anterior-Posterior (AP) CXR images from DICOM to JPG format. Specifically, pixel values in the DICOM format were normalized to a range of {[}0, 255{]}. If necessary, all pixels were inverted to ensure that the air in the image appeared white, and the area outside the patient's body appeared black. Following that, we performed histogram equalization to enhance the image contrast. Finally, the processed image was saved in JPG format with a quality factor of 95.

All images were subsequently resized to \(256 \times 256 \times 3\) using PyTorch\textquotesingle s default bilinear interpolation and center-cropped to \(224 \times 224 \times 3\). A stochastic image augmentation was randomly applied to transform a CXR into an augmented view. In this work, we sequentially applied two simple augmentation operations: (1) random rotation between \(0^{{^\circ}}\) and \(10^{{^\circ}}\), and (2) random flipping.

The proposed method is not exclusive to specific deep learning models, and Densenet-121\cite{Huang2017-xq} showed good performance in classification on NIH-CXR in the previous study.\cite{Rajpurkar2017-se} Therefore, we used a Densenet-121 architecture pretrained on CheXpert\cite{Irvin2019-eg} in this study. The last layer of the model is firstly substituted by a single-layer perception with an output dimension of 128 (backbone \(+\) contrastive head). We then fine-tuned the entire network for the subsequential tasks. Adam optimizer\cite{Kingma2015-kg} with a learning rate of 0.0001 was used for contrastive learning. We set the temperature to 0.05 and trained the model for 10 epochs. After that, we replaced the output layer with the classification output layer (backbone \(+\) prediction head) and fine-tuned the model for 1 epoch. The experiments were conducted on an Intel Core i9-9960 X 16-core processor and an NVIDIA Quadro RTX 6000 GPU. The models were implemented using PyTorch. The code is available at \url{https://github.com/bionlplab/CXRFairness}.

For the MIDRC dataset, we randomly split the entire dataset at the patient level. We designated one group (20\% of the total subjects) as the hold-out test set and used the remaining portion as the training and validation sets. For the NIH-CXR dataset, we used the official training, validation, and testing split.

We evaluated our methods on all subgroups across age, gender, and race, and compared our result with four baselines: Empirical Risk Minimization (ERM),\cite{Vapnik1991-ro} balanced ERM,\cite{Zhang2022-rr} Adversarial,\cite{Wadsworth2018-ob} and supervised contrastive learning (SCL).\cite{Khosla2020-kq} In this study, these four baselines are based on DenseNet-121 pretrained on CheXpert,\cite{Irvin2019-eg} which we refer to as DenseNet-121, Balance DenseNet-121, ADV, and SCL. DenseNet-121 is an original algorithm that does not consider the bias problem, and our proposed algorithm uses DenseNet-121 as its backbone. Data resampling is a commonly used data pre-processing technique for reducing bias in subgroups, so we employed Balanced DenseNet-121 as one of the baselines. In this study, we resampled the subgroups with fewer samples to ensure that the number of samples in all subgroups was the same. ADV is a widely used in-processing method derived from the domain adaptation field, which treats the sensitive attribute as a domain-specific label and attempts to use only domain-irrelevant features for the target task. SCL is a general contrastive learning approach without label definitions related to demographic information, which we use to demonstrate the effectiveness of the proposed method.

\subsection{Statistical Analysis}

We used 200 bootstrap samples to obtain a distribution of the \dmauc and reported 95\% confidence intervals. For each bootstrap iteration, we sampled \(n\) images with replacements from the test set of \(n\) images. To compare the difference in \dmauc between the proposed model and baseline across all subgroups, we conducted a paired t-test. Statistical analysis was conducted using SciPy 1.7.1 with statistical significance defined as \(p < 0.05\).

\section{Results}

\subsection{Study Participants}

Table \ref{tab:characteristics} lists the patient characteristics for both datasets. For the MIDRC dataset, the training set includes 22,237 patients (median age, 59 years; interquartile range {[}IQR{]}: 44-71 years; 11,257 {[}51\%{]} men), and the test set includes 5,559 patients (median age, 59 years; interquartile range {[}IQR{]}: 25-75 years; 2,819 {[}51\%{]} men). For the NIH-CXR dataset, the training set includes 28,008 patients (median age, 48 years; interquartile range {[}IQR{]}: 34-59 years; 15,073 {[}54\%{]} men), and the test set includes 2,797 patients (median age, 49 years; interquartile range {[}IQR{]}: 34-59 years; 1,557 {[}54\%{]} men).
\begin{table}
\centering
\caption{Patent characteristics for both datasets}
\label{tab:characteristics}
\begin{threeparttable}
\footnotesize
\begin{tabular}{llrr}
\toprule
Dataset & Characteristic             & Training   set & Test   set   \\
\midrule
MIDRC   \\
        & Age,   yrs (interquartile) & 59   (44-71)   & 59   (25-75) \\
        & Sex                        &                &              \\
        & \hspace{2em}Male, n (\%)               & 11,257   (51)  & 2,819   (51) \\
        & \hspace{2em}Female, n (\%)             & 10,980   (49)  & 2,740   (49) \\
        & Race                       &                &              \\
        & \hspace{2em}White, n (\%)              & 12,002   (54)  & 2,998   (54) \\
        & \hspace{2em}Black, n (\%)              & 7,444   (33)   & 1,912   (34) \\
        & \hspace{2em}Other, n (\%)              & 2,791   (13)   & 649   (12)   \\
\midrule
NIH-CXR \\
        & Age,   yrs                 & 48   (34-59)   & 49   (34-59) \\
        & Sex                        &                &              \\
        & \hspace{2em}Male, n (\%)*              & 15,073   (54)  & 1,557   (56) \\
        & \hspace{2em}Female, n (\%)             & 12,935   (46)  & 1,240   (44)\\
\bottomrule
\end{tabular}
\begin{tablenotes}
\item The category Other includes American Indian or Alaska Native, Asian, Native Hawaiian or other Pacific Islander, and Other.
\end{tablenotes}
\end{threeparttable}
\end{table}

Table \ref{tab:subgroup} presents the subgroup information of the datasets at the image level. Our study focused on training image-based classifiers for disease detection and evaluating the model's performance on the subgroups based on sex, age, and race for the MIDRC dataset, and sex and age for the NIH-CXR dataset. Due to the different age characteristics between these two datasets (Table \ref{tab:characteristics}), we set the age groups for MIDRC and NIH datasets differently.

\begin{table}
\centering
\caption{The subgroup information of two datasets at the image level: MIDRC\cite{Lakhani2023-lp} and Chest X-ray14\cite{Wang2017-ChestX}}
\label{tab:subgroup}
\begin{threeparttable}
\footnotesize
\begin{tabular}{llrrrr}
\toprule
Dataset & Characteristic & \multicolumn{2}{c}{Training set} & \multicolumn{2}{c}{Test set} \\
\cmidrule(lr){3-4}\cmidrule(lr){5-6}
 &  & Positive & Total & Positive & Total \\
\midrule
MIDRC \\
 & Images,   n & 31,434 & 62,178 & 7,935 & 15,709 \\
 & Age &  &  &  &  \\
 & \hspace{2em}$<$ 75 yrs & 21,213 & 52,427 & 6,970 & 13,115 \\
 & \hspace{2em}$\ge$ 75 yrs & 4,566 & 9,751 & 965 & 2,594 \\
 & Sex &  &  &  &  \\
 & \hspace{2em}Male & 17,991 & 35,081 & 4,404 & 8,799 \\
 & \hspace{2em}Female & 13,443 & 27,097 & 3,531 & 6,910 \\
 & Race &  &  &  &  \\
 & \hspace{2em}White & 11,616 & 30,667 & 2,739 & 7,790 \\
 & \hspace{2em}Black & 16,836 & 24,104 & 4,456 & 6,135 \\
 & \hspace{2em}Other & 2,982 & 7,407 & 7,40 & 1,784 \\
\midrule
NIH-CXR \\
 & Images,   n & 43,021 & 86,524 & 9,671 & 25,596 \\
 & Age &  &  &  &  \\
 & \hspace{2em}$<$ 60 yrs & 30,933 & 66,048 & 7,579 & 19,634 \\
 & \hspace{2em}$\ge$ 60 yrs & 12,088 & 20,476 & 2,092 & 5,962 \\
 & Sex &  &  &  &  \\
 & \hspace{2em}Male & 24,409 & 48,858 & 5,595 & 14,882 \\
 & \hspace{2em}Female & 18,612 & 38,066 & 4,076 & 10,714\\
\bottomrule
\end{tabular}
\begin{tablenotes}
\item The category Other includes American Indian or Alaska Native, Asian, Native Hawaiian or other Pacific Islander, and Other.
\end{tablenotes}
\end{threeparttable}
\end{table}

\subsection{Data investigation}

We first employed logistic regression to analyze the association between demographic information (age, sex, and race) and the prevalence of COVID-19 on the MIDRC dataset. Age, sex, and race were used as predictors and compared with the reference group (e.g., individuals under 75 years versus those aged 75 years and above, male versus female, Black versus White, and other racial individuals versus White). Odds ratios (ORs) larger than 1 indicated that the comparison groups had higher rates of antecedent compared to the reference group. $p<0.05$ is considered significant. When comparing rates, 95\% confidence intervals (CI) were calculated.

\begin{figure}
    \centering
    \includegraphics[width=.5\linewidth]{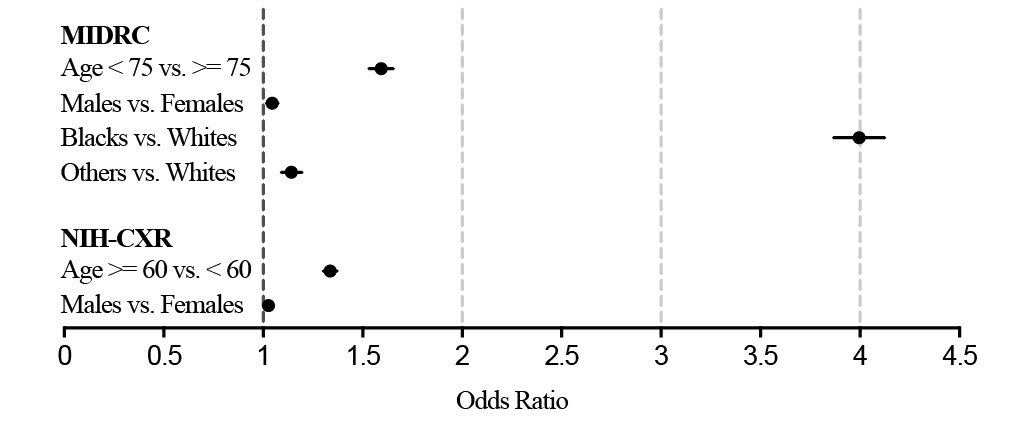}
    \caption{Forest plot of relative odds (95\% confidence intervals) of COVID-19 (MIDRC) and thorax abnormality (NIH-CXR) associated with age, sex, and race.}
    \label{fig:relative}
\end{figure}

As shown in Figure \ref{fig:relative}, COVID-19 is associated with individuals under 75 years (OR = 1.59, 95\% CI = 1.53-1.66), males (OR = 1.04, 95\% CI = 1.02-1.08), Black (OR = 4.00, 95\% CI = 3.87-4.13), and other racial (OR = 1.14, 95\% CI = 1.09-1.20).

We also examined the association between demographic factors (age and sex) and thorax abnormality on the NIH-CXR dataset. Figure \ref{fig:relative} shows that thorax abnormality is associated with individuals aged 60 years or older (OR = 1.34, 95\% CI = 1.30-1.37) and males (OR = 1.03, 95\% CI: 1.00-1.05).

\subsection{Model Fairness Comparisons in MIDRC Dataset}

\begin{figure}
    \centering
    \includegraphics[width=.6\linewidth]{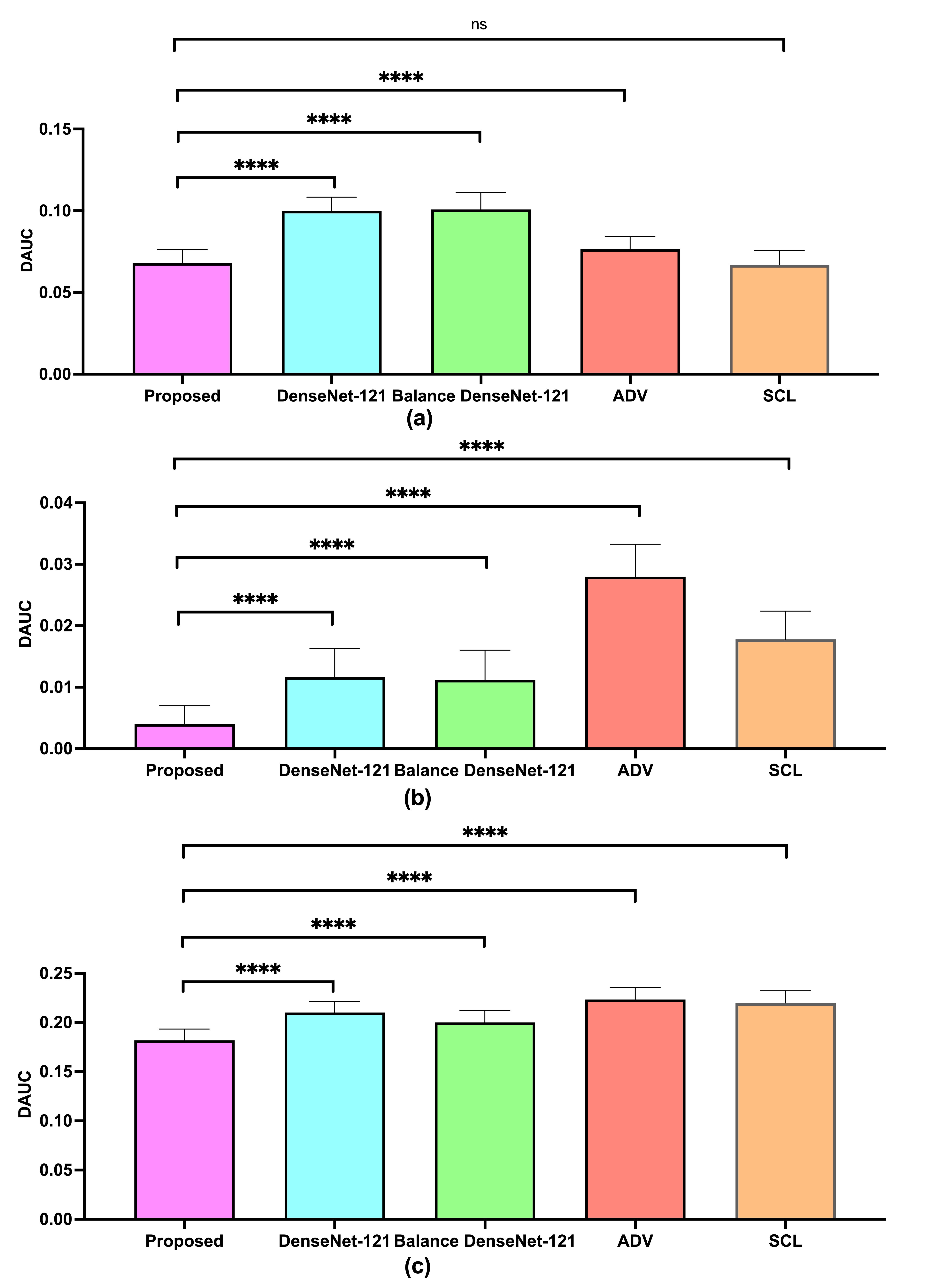}
    \caption{\dmauc across subgroups of sex (a), age (b), and race (c) in COVID-19 detection on the MIDRC dataset.  The results are averaged over 200 times bootstrap experiment. ****: p-value $\leq$ 0.0001. Balance DenseNet-121 – DenseNet-121 with balanced empirical risk minimization\cite{Vapnik1991-ro} ADV - Adversarial.\cite{Wadsworth2018-ob} SCL - supervised contrastive learning.\cite{Khosla2020-kq}}
    \label{fig:midrc result}
\end{figure}

Figure \ref{fig:midrc result} shows that our proposed method produced significantly smaller \dmauc across all demographics in comparison to the baselines. This demonstrates the effectiveness of our approach in reducing bias and promoting fairness in AI models for COVID-19 diagnosis on CXRs.

Table \ref{tab:auc midrc} presents a detailed performance comparison of various methods for COVID-19 diagnosis based on sex, race, and age. Individuals in subgroups with lower AUC values are at a higher risk of being misdiagnosed compared to their counterparts.
\begin{table}
\caption{The AUC, marginal AUC, and marginal AUC difference (\dmauc) of baseline and proposed model for COVID-19 diagnosis in the MIDRC dataset. Balance DenseNet-121 – DenseNet-121 with balanced empirical risk minimization.\cite{Vapnik1991-ro} ADV - Adversarial.\cite{Wadsworth2018-ob} SCL - supervised contrastive learning.\cite{Khosla2020-kq}}
\label{tab:auc midrc}
\resizebox{\linewidth}{!}{
\begin{tabular}{lccccc}
\toprule
& DenseNet-121 & Balance DenseNet-121 & ADV & SCL & Proposed\\
\midrule
Sex &  &  &  &  &  \\
\hspace{2em}Overall AUC & 0.8167 (0.8162-0.8171) & 0.8110 (0.8106-0.8116) & 0.8060 (0.8056-0.8065) & 0.8090 (0.8085-0.8095) & 0.8085 (0.8080-0.8090)\\
\hspace{2em}Male & 0.8218 (0.8213-0.8224) & 0.8161 (0.8155-0.8166) & 0.8185 (0.8180-0.8190) & 0.8169 (0.8164-0.8175) & 0.8088 (0.8083-0.8094) \\
\hspace{2em}Female & 0.8102 (0.8096-0.8108) & 0.8049 (0.8043-0.8056) & 0.7905 (0.7899-0.7916) & 0.7991 (0.7985-0.7998) & 0.8081 (0.8075-0.8073) \\
\hspace{2em}\dmauc & 0.0116 (0.0110-0.0123) & 0.0112 (0.0105-0.0119) & 0.0280 (0.0273-0.0287) & 0.0179 (0.0173-0.0184) & 0.0037 (0.0032-0.0041) \\
Race &  &  &  &  &  \\
\hspace{2em}Overall AUC & 0.8167 (0.8162-0.8171) & 0.8051 (0.8048-0.8056) & 0.8157 (0.8152-0.8161) & 0.8106 (0.8101-0.8111) & 0.7918 (0.7913-0.7923)\\
\hspace{2em}White & 0.7583 (0.7577-0.7590) & 0.7715 (0.7708-0.7721) & 0.7568 (0.7561-0.7575) & 0.7551 (0.7544-0.7559) & 0.7515 (0.7507-0.7522) \\
\hspace{2em}Black & 0.8775 (0.8770-0.8779) & 0.8515 (0.8511-0.8519) & 0.8787 (0.8782-0.8791) & 0.8712 (0.8708-0.8716) & 0.8391 (0.8386-0.8396) \\
\hspace{2em}Other & 0.6672 (0.6657-0.6688) & 0.6513 (0.6496-0.6530) & 0.6552 (0.6535-0.6569) & 0.6513 (0.6496-0.6530) & 0.6572 (0.6556-0.6589) \\
\hspace{2em}\dmauc & 0.2103 (0.2087-0.2118) & 0.2002 (0.1985-0.2019) & 0.2235 (0.2217-0.2251) & 0.2199 (0.2182-0.2216) & 0.1819 (0.1802-0.1835) \\
Age &  &  &  &  &  \\
\hspace{2em}Overall AUC & 0.8167 (0.8162-0.8171) & 0.8020 (0.8016-0.8025) & 0.8134 (0.8130-0.8139) & 0.8077 (0.8072-0.8082) & 0.8011 (0.8006-0.8015) \\
\hspace{2em}$<$ 75 yrs & 0.8288 (0.8284-0.8293) & 0.8143 (0.8139-0.8148) & 0.8227 (0.8223-0.8232) & 0.8158 (0.8153-0.8163) & 0.8094 (0.8089-0.8098) \\
\hspace{2em}$\ge$ 75 yrs & 0.7289 (0.7277-0.7300) & 0.7135 (0.7121-0.7149) & 0.7463 (0.7452-0.7473) & 0.7488 (0.7476-0.7501) & 0.7414 (0.7403-0.7425) \\
\hspace{2em}\dmauc & 0.0999 (0.0988-0.1011) & 0.1008 (0.0994-0.1023) & 0.0764 (0.0754-0.0776) & 0.0670 (0.0658-0.0682) & 0.0680 (0.0669-0.0692) \\
\bottomrule
\end{tabular}
}
\end{table}

Specifically, compared to DenseNet-121, the \dmauc obtained by the proposed method decreased from 0.0116 (95\% CI, 0.0110-0.0123) to 0.0040 (95\% CI, 0.0036-0.0044) for sex, with female individuals showing lower marginal AUC values than their male counterparts. For the race subgroup, the \dmauc obtained by the proposed method decreased from 0.2102 (95\% CI, 0.2087-0.2118) to 0.1818 (95\% CI, 0.1802-0.1835) compared to DenseNet-121. The "other" group showed lower marginal AUC values than the White and Black groups. Similarly, for the age subgroup, the \dmauc values decreased from 0.0999 (95\% CI, 0.0988-0.1011) to 0.0686 (95\% CI, 0.0669-0.0692) compared with DenseNet-121, with individuals below 75 years of age displaying lower marginal AUC values than their counterparts. The individuals in subgroups with lower marginal AUC are more likely to be misdiagnosed than their counterparts.

Supplementary Tables \ref{tab:tpr midrc}, \ref{tab:fpr midrc}, and \ref{tab:bs midrc} present TPR and \dtpr, FPR and \dfpr, and BS and \dbs of various methods for COVID-19 diagnosis based on sex, race, and age, respectively. The proposed method demonstrated comparable TPR in sex and race to DenseNet-121, with lower \dtpr in sex, race, and age. Moreover, the proposed method exhibited lower FPR in age and reduced \dfpr in race and age compared to DenseNet-121. Additionally, the proposed method generated lower \dbs in sex and age, and lower BS in age.

\subsection{Model Fairness Comparisons in NIH-CXR Dataset}

\begin{figure}
    \centering
    \includegraphics[width=.6\linewidth]{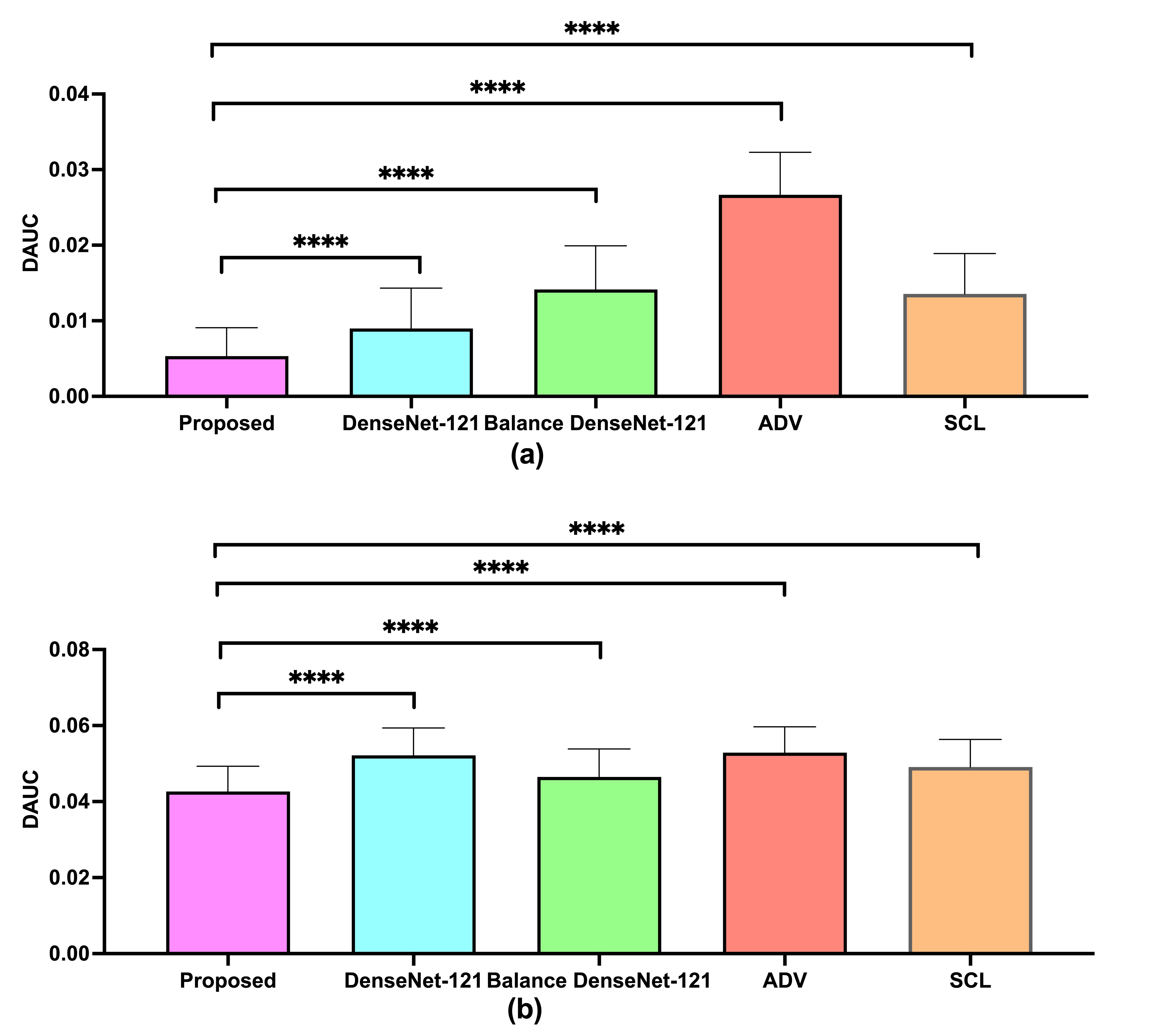}
    \caption{\dmauc across subgroups of sex (a) and age (b) in the thorax abnormality detection on the NIH-CXR dataset. The results are averaged over 200 times bootstrap experiment. ****: p-value $\leq$ 0.0001. Balance DenseNet-121 – DenseNet-121 with balanced empirical risk minimization.\cite{Vapnik1991-ro} ADV - Adversarial.\cite{Wadsworth2018-ob} SCL - supervised contrastive learning.\cite{Khosla2020-kq}}
    \label{fig:nih}
\end{figure}

For diagnosing thorax abnormalities on the NIH-CXR dataset, Figure \ref{fig:nih} shows that our proposed method also produced significantly smaller \dmauc across all demographics in comparison to the baselines.

Table \ref{tab:auc nih} further presents a detailed analysis of the results. The proposed method achieved a lower \dmauc compared to the baselines for all demographic groups. In the race subgroup analysis, the \dmauc obtained by the proposed method decreased from 0.0090 (95\%, 0.0082-0.0097) to 0.0053 (95\% CI, 0.0048-0.0059) compared to DenseNet-121. The proposed model performed similarly for male individuals as their counterparts in the race subgroup analysis, while the baselines generated lower AUC for male individuals. In the age subgroup, the \dmauc obtained by the proposed method decreased from 0.0512 (95\%CI, 0.0512-0.0532) to 0.0427 (95\%CI,0.0417-0.0436) compared to the DenseNet-121. Individuals over 60 years of age had lower AUCs than their counterparts.
\begin{table}
\caption{The AUC, marginal AUC, and marginal AUC difference (\dmauc) of baseline and proposed model for thorax abnormalities diagnosis in the NIH-CXR dataset. Balance DenseNet-121 – DenseNet-121 with balanced empirical risk minimization.\cite{Vapnik1991-ro} ADV - Adversarial.\cite{Wadsworth2018-ob} SCL - supervised contrastive learning.\cite{Khosla2020-kq}}
\label{tab:auc nih}
\resizebox{\linewidth}{!}{
\begin{tabular}{lccccc}
\toprule
& DenseNet-121 & Balance DenseNet-121 & ADV & SCL & Proposed\\
\midrule
Sex &  &  &  &  &  \\
\hspace{2em}Overall AUC&  0.7354 (0.7349-0.7369) & 0.7322 (0.7317-0.7327) & 0.7108 (0.7102-0.7112) & 0.7306 (0.7301-0.7310) & 0.7341 (0.7336-0.7346) \\
\hspace{2em}Male & 0.7317 (0.7311-0.7324) & 0.7263 (0.7256-0.7268) & 0.6995 (0.6988-0.7001) & 0.7249 (0.7243-0.7255) & 0.7355 (0.7349-0.7361) \\
\hspace{2em}Female & 0.7404 (0.7398-0.7411) & 0.7404 (0.7397-0.7410) & 0.7261 (0.7255-0.7268) & 0.7384 (0.7378-0.7390) & 0.7322 (0.7316-0.7328) \\
\hspace{2em}\dmauc &  0.0090 (0.0082-0.0097) & 0.0142 (0.0133-0.0150) & 0.0267 (0.0259-0.0275) & 0.0135 (0.0128-0.0143) & 0.0053 (0.0048-0.0059) \\
Age &  &  &  &  &  \\
\hspace{2em}Overall AUC&  0.7354 (0.7349-0.7369) & 0.7243 (0.7243 – 0.7252) & 0.7071 (0.7066-0.7076) & 0.7306 (0.7302-0.7311) & 0.7277 (0.7272-0.7282)  \\
\hspace{2em}$<$ 60 yrs & 0.7467 (0.7461-0.7472) & 0.7348 (0.7343-0.7353) & 0.7185 (0.7180-0.7191) & 0.7413 (0.7407-0.7418) & 0.7369 (0.7364-0.7374) \\
\hspace{2em}$\ge$ 60 yrs & 0.6945 (0.6936-0.6954) & 0.6883 (0.6874-0.6893) & 0.6656 (0.6648-0.6665) & 0.6922 (0.6913-0.6931) & 0.6943 (0.6935-0.6951) \\
\hspace{2em}\dmauc &  0.0512 (0.0512-0.0532) & 0.0465 (0.0455-0.0475) & 0.0529 (0.0520-0.0538) & 0.0491 (0.0481-0.0501) & 0.0427 (0.0417-0.0436)  \\
\bottomrule
\end{tabular}
}
\end{table}

Supplementary Tables \ref{tab:tpr nih}, \ref{tab:fpr nih}, and \ref{tab:bs nih} list TPR and \dtpr, FPR and \dfpr, and BS and \dbs of various methods for diagnosing thorax abnormalities on the NIH-CXR dataset across sex and age, respectively. The proposed method achieved higher TPR on sex and age than DenseNet-121. Additionally, the proposed method exhibited lower \dfpr on sex than DenseNet-121. Furthermore, the proposed method generated comparable BS and lower \dbs in sex and age than DenseNet-121.

\subsection{External Validation}

To further evaluate the proposed method, we utilized both DenseNet-121 and the proposed model trained on the NIH-CXR dataset, testing them on the MIMIC-CXR test set. The details of the results are presented in Supplementary Table \ref{tab:auc cross}. For external validation, our proposed method achieved a higher AUC for both sex and age, as well as a lower \dmauc for sex.

\section{Discussion}

In this study, we proposed a method leveraging supervised contrastive learning to reduce bias in AI models for CXR image diagnosis across different groups. The proposed model was evaluated using two large-scale CXR datasets. We observed systematic model biases in subgroups across all settings. This highlights the importance of addressing biases in AI models to ensure fair and accurate diagnoses across all demographic subgroups. We have the following observations for further discussion.

First, the logistic regression analysis revealed the existence of bias in the dataset across different demographics. For example, the results indicate that individuals aged 60 years or older are more likely to show abnormalities compared to those younger than 60 years (OR = 1.34, 95\% CI = 1.30-1.37) in the NIH-CXR dataset.

Second, our proposed method effectively improves the fairness of CXR diagnoses by utilizing supervised contrastive learning to obtain fair image embeddings that retain label information. In contrast to state-of-the-art models, we modified the definitions of positive and negative samples to enable the network to capture more label information and less group information. Our proposed method generated smaller \dmauc for both datasets across all demographics when compared to the baselines. Additionally, the proposed method consistently maintains overall performance. We conduct quantitative analysis using the metric of relative change. Supplementary Table \ref{tab:rel abs} shows that the relative change in AUC and mAUC remains consistently within 2\%, while the relative change in \dmauc ranges from 13.5\% to 68.10\%. The results suggest that our proposed method can successfully reduce bias (\dmauc) without significantly compromising AUC and mAUC.

Third, this study highlights the impact of data imbalance on the bias of AI models. The overrepresentation of prevalent patients in certain subgroups can lead to biased models, as evidenced by our findings (Table 2). For example, in the MIDRC dataset, the prevalence of COVID-19 is significantly higher among Black individuals compared to their White counterparts (70.02\% vs. 37.33\%). This overrepresentation can result in biased models trained on this dataset. Additionally, even when subgroups have similar prevalence, the sample size can still introduce bias. For instance, in the MIDRC dataset, the number of White, Black, and ``Other'' individuals are 38,457, 30,239, and 9,191, respectively. Although the COVID-19 prevalence is almost the same for the ``Other'' and White individuals (40.50\% vs 37.33\%), the former group, which had the smallest sample size among the racial subgroups, obtained the lowest AUC value. Similar phenomena were observed in the age subpopulations for thorax disease detection in the NIH-CXR dataset and for COVID-19 detection in the MIDRC dataset.

Fourth, data resampling is a commonly used data pre-processing technique for reducing bias in subgroups, but our findings suggest that it may not always be effective. Specifically, our results in Table 3 and Table 4 show that the Balance DenseNet-121 model could not reduce bias for sex and age on the MIDRC dataset, or bias for sex on the NIH-CXR dataset, compared with the DenseNet-121 model. In this study, we only employed one resampling method to ensure an equal sample size across subgroups. However, future research could explore additional resampling methods to determine their effectiveness.

Fifth, ADV is widely used as an in-processing method to improve group fairness, but our results suggest that it may not always be effective. Specifically, the results in Table 3 and Table 4 show that the ADV model could only reduce bias related to age in the MIDRC dataset when compared with the DenseNet-121 model.

Finally, SCL is a general contrastive learning approach without label definitions related to demographic information. The experiments conducted with SCL can be regarded as an ablation study to demonstrate that our proposed method considers demographic information to form positive and negative samples for learning image feature embeddings to improve group fairness. The results in Table 3 and Table 4 show that our proposed method effectively improve group fairness.

While this study assessed the fairness of binarized models, one limitation is that it did not examine the calibration of predicted probabilities. As a result, there is a possibility of overconfidence or underconfidence in certain cases. To address this limitation, future research should investigate the relationship between calibration and bias in disease detection and develop effective methods to reduce calibration bias. Moreover, expanding the proposed method to include continuous attributes, in addition to discrete groups, and multi-class settings, would increase its applicability. Additionally, this study aims to enhance the fairness of automated Chest X-ray diagnosis through contrastive learning. We utilized two extensive Chest X-ray datasets to showcase the effectiveness of the proposed method, with both datasets focusing on thoracic diseases. In the future, we plan to extend the application of our method to other diseases and imaging modalities and test on more models. Furthermore, the data does not provide the comorbidity history of the patients.

In summary, this study introduces an effective AI model that reduces bias toward subgroups in CXR diagnosis. Notably, this represents the first attempt to address bias in deep learning for COVID-19 diagnosis. Our proposed approach uses supervised contrastive learning as a pretraining method to obtain fair image embeddings. Unlike previous supervised contrastive methods, our approach uses images with the same label but from different protected groups as positive samples and images with different labels but from the same protected group as negative samples for each anchor image in a minibatch. This allows the network to capture more label information and less group information during pretraining. The backbone of the model is fine-tuned in the downstream task. We developed and evaluated the proposed method using two large multi-institutional datasets, which demonstrated its effectiveness in reducing bias. Therefore, the proposed method is suitable for clinical practice and can help alleviate concerns regarding disparities generated by AI models.

\section*{Disclosures of Conflicts of Interest}

No conflicting relationship exists for any author.

\section*{Acknowledgment}

This work was supported by the National Institutes of Health (NIH), National Eye Institute under Award No. R21EY035296 (Y.P., F.W., S.H.VT), National Science Foundation CAREER Award No. 2145640 (Y.P.), Amazon Research Award (Y.P., G.S.), and NIH Artificial Intelligence/Machine Learning Consortium to Advance Health Equity and Researcher Diversity (AIM-AHEAD) Program Award No. OT2OD032581 (Y.P., Y.D.).

\bibliographystyle{unsrtnat}
\bibliography{ref}

\newpage
\appendix
\setcounter{table}{0}
\setcounter{figure}{0}
\renewcommand\figurename{Supplementary Figure} 
\renewcommand\tablename{Supplementary Table}

\section*{Supplementary materials}

\begin{table}[!hbpt]
\caption{The TPR, and \dtpr of baseline and proposed model for COVID-19 diagnosis in the MIDRC dataset. Balance DenseNet-121 – DenseNet-121 with balanced empirical risk minimization.\cite{Vapnik1991-ro} ADV - Adversarial.\cite{Wadsworth2018-ob} SCL - supervised contrastive learning.\cite{Khosla2020-kq}}
\label{tab:tpr midrc}
\resizebox{\linewidth}{!}{
\begin{tabular}{lccccc}
\toprule
& DenseNet-121 & Balance DenseNet-121 & ADV & SCL & Proposed\\
\midrule
Sex &  &  &  &  & \\
\hspace{2em}Overall & 0.7786 (0.7779-0.7792) & 0.7042 (0.7035-0.7050) & 0.7503 (0.7496-0.7510) & 0.7274 (0.7268-0.7281) & 0.7767 (0.7761-0.7774)\\
\hspace{2em}Male & 0.7822 (0.7813-0.7831) & 0.7063 (0.7053-0.7073) & 0.7668 (0.7659-0.7678) & 0.7389 (0.7380-0.7398) & 0.7765 (0.7757-0.7774)\\
\hspace{2em}Female & 0.7740 (0.7729-0.7751) & 0.7016 (0.7006-0.7027) & 0.7296 (0.7287-0.7306) & 0.7131 (0.7120-0.7141) & 0.7770 (0.7760-0.7780)\\
\hspace{2em}\dtpr & 0.0104 (0.0093-0.0114) & 0.0086 (0.0077-0.0095) & 0.0372 (0.0359-0.0385) & 0.0258 (0.0244-0.0272) & 0.0074 (0.0066-0.0082)\\
\midrule
Race &  &  &  &  & \\
\hspace{2em}Overall & 0.7786 (0.7779-0.7792) & 0.6936 (0.6929-0.6934) & 0.7513 (0.7506-0.7520) & 0.6952 (0.6945-0.6959) & 0.7794 (0.7788-0.7801)\\
\hspace{2em}White & 0.6809 (0.6796-0.6821) & 0.6290 (0.6257-0.6281) & 0.6470 (0.6457-0.6483) & 0.5799 (0.5787-0.5811) & 0.7221 (0.7210-0.7235)\\
\hspace{2em}Black & 0.8713 (0.8706-0.8720) & 0.7661 (0.7652-0.7669) & 0.8517 (0.8509-0.8524) & 0.8051 (0.8043-0.8059) & 0.8419 (0.8412-0.8427)\\
\hspace{2em}Other & 0.5813 (0.5788-0.5838) & 0.5039 (0.5012-0.5066) & 0.5329 (0.5303-0.5355) & 0.4594 (0.4567-0.4622) & 0.6147 (0.6119-0.6174)\\
\hspace{2em}\dtpr & 0.2900 (0.2874-0.2926) & 0.2622 (0.2594-0.2650) & 0.3188 (0.3161-0.3214) & 0.3456 (0.3428-0.3485) & 0.2273 (0.2245-0.2301)\\
\midrule
Age &  &  &  &  & \\
\hspace{2em}Overall & 0.7786 (0.7779-0.7792) & 0.7377 (0.7370-0.7385) & 0.6851 (0.6844-0.6859) & 0.7718 (0.7711-0.7725) & 0.6699 (0.6691-0.6706)\\
\hspace{2em}$<$ 75 yrs & 0.8012 (0.8005-0.8019) & 0.7593 (0.7585-0.7601) & 0.7076 (0.7069-0.7084) & 0.7850 (0.7843-0.7857) & 0.6859 (0.6851-0.6867)\\
\hspace{2em}$\geq$ 75 yrs & 0.6145 (0.6122-0.6167) & 0.5813 (0.5790-0.5837) & 0.5219 (0.5195-0.5244) & 0.6758 (0.6753-0.6780) & 0.5535 (0.5512-0.5558)\\
\hspace{2em}\dtpr & 0.1867 (0.1844-0.1891) & 0.1780 (0.1755-0.1805) & 0.1857 (0.1831-0.1883) & 0.1092 (0.1069-0.1116) & 0.1324 (0.1299-0.1349)\\
\bottomrule
\end{tabular}
}
\end{table}

\newpage

\begin{table}[!hbpt]
\caption{The FPR, and \dfpr of baseline and proposed model for COVID-19 diagnosis in the MIDRC dataset. Balance DenseNet-121 – DenseNet-121 with balanced empirical risk minimization .\cite{Vapnik1991-ro} ADV - Adversarial.\cite{Wadsworth2018-ob} SCL - supervised contrastive learning.\cite{Khosla2020-kq}}
\label{tab:fpr midrc}
\resizebox{\linewidth}{!}{
\begin{tabular}{lccccc}
\toprule
& DenseNet-121 & Balance DenseNet-121 & ADV & SCL & Proposed\\
\midrule
Sex &  &  &  &  & \\
\hspace{2em}Overall & 0.2868 (0.2861-0.2875) & 0.2245 (0.2239-0.2252) & 0.2784 (0.2777-0.2791) & 0.2529 (0.2522-0.2536) & 0.3031 (0.3024-0.3039)\\
\hspace{2em}Male & 0.2836 (0.2826-0.2846) & 0.2187 (0.2178-0.2196) & 0.2841 (0.2831-0.2851) & 0.2548 (0.2538-0.2557) & 0.2838 (0.2828-0.2848)\\
\hspace{2em}Female & 0.2909 (0.2898-0.2919) & 0.2321 (0.2311-0.2331) & 0.2709 (0.2699-0.2720) & 0.2505 (0.2495-0.2516) & 0.3283 (0.3271-0.3294)\\
\hspace{2em}\dfpr & 0.0100 (0.0089-0.0111) & 0.0143 (0.0130-0.0155) & 0.0145 (0.0132-0.0157) & 0.0089 (0.0080-0.0097) & 0.0445 (0.0429-0.0460)\\
\midrule
Race &  &  &  &  & \\
\hspace{2em}Overall & 0.2868 (0.2861-0.2875) & 0.2140 (0.2134-0.2146) & 0.2640 (0.2633-0.2647) & 0.2236 (0.2229-0.2243) & 0.3500 (0.3493-0.3508)\\
\hspace{2em}White & 0.2219 (0.2211-0.2227) & 0.1728 (0.1721-0.1735) & 0.1985 (0.1977-0.1992) & 0.1697 (0.1689-0.1705) & 0.2863 (0.2855-0.2872)\\
\hspace{2em}Black & 0.5448 (0.5430-0.5466) & 0.3931 (0.3915-0.3948) & 0.5203 (0.5187-0.5220) & 0.4431 (0.4413-0.4449) & 0.5806 (0.5790-0.5821)\\
\hspace{2em}Other & 0.1857 (0.1840-0.1873) & 0.1250 (0.1235-0.1265) & 0.1689 (0.1673-0.1706) & 0.1313 (0.1300-0.1326) & 0.2874 (0.2855-0.2893)\\
\hspace{2em}\dfpr & 0.3592 (0.3566-0.3617) & 0.2622 (0.2594-0.2650) & 0.3514 (0.3491-0.3537) & 0.3118 (0.3095-0.3141) & 0.2931 (0.2906-0.2956)\\
\midrule
Age &  &  &  &  & \\
\hspace{2em}Overall & 0.2868 (0.2861-0.2875) & 0.2671 (0.2664-0.2678) & 0.2066 (0.2060-0.2072) & 0.3098 (0.3091-0.3105) & 0.2198 (0.2192-0.2204)\\
\hspace{2em}$<$ 75 yrs & 0.3107 (0.3099-0.3115) & 0.2949 (0.2941-0.2957) & 0.2250 (0.2243-0.2257) & 0.3369 (0.3361-0.3377) & 0.2416 (0.2409-0.2423)\\
\hspace{2em}$\geq$ 75 yrs & 0.1968 (0.1953-0.1982) & 0.1623 (0.1610-0.1636) & 0.1373 (0.1361-0.1384) & 0.2077 (0.2064-0.2090) & 0.1376 (0.1364-0.1388)\\
\hspace{2em}\dfpr & 0.1139 (0.0988-0.1011) & 0.1326 (0.1311-0.1340) & 0.0878 (0.0864-0.0891) & 0.1292 (0.1277-0.1308) & 0.1040 (0.1027-0.1054)\\
\bottomrule
\end{tabular}
}
\end{table}

\newpage

\begin{table}[!hbpt]
\caption{The brier scores (BS), and \dbs of baseline and proposed model for COVID-19 diagnosis in the MIDRC dataset. Balance DenseNet-121 – DenseNet-121 with balanced empirical risk minimization.\cite{Vapnik1991-ro} ADV - Adversarial.\cite{Wadsworth2018-ob} SCL - supervised contrastive learning.\cite{Khosla2020-kq}}
\label{tab:bs midrc}
\resizebox{\linewidth}{!}{
\begin{tabular}{lccccc}
\toprule
& DenseNet-121 & Balance DenseNet-121 & ADV & SCL & Proposed\\
\midrule
Sex &  &  &  &  & \\
\hspace{2em}Overall & 0.1766 (0.1763-0.1769) & 0.1810 (0.1807-0.1812) & 0.1819 (0.1816-0.1821) & 0.1793 (0.1790-0.1795) & 0.1804 (0.1802-0.1807)\\
\hspace{2em}Male & 0.1802 (0.1799-0.1805) & 0.1722 (0.1719-0.1740) & 0.1797 (0.1794-0.1800) & 0.1755 (0.1752-0.1758) & 0.1876 (0.1873-0.1879)\\
\hspace{2em}Female & 0.1852 (0.1849-0.1855) & 0.1737 (0.1734-0.1740) & 0.1896 (0.1893-0.1899) & 0.1810 (0.1808-0.1813) & 0.1901 (0.1898-0.1904)\\
\hspace{2em}\dbs & 0.0050 (0.0048-0.0052) & 0.0019 (0.0017-0.0021) & 0.0099 (0.0096-0.0101) & 0.0056 (0.0054-0.0058) & 0.0026 (0.0024-0.0028)\\
\midrule
Race &  &  &  &  & \\
\hspace{2em}Overall & 0.1767 (0.1763-0.1769) & 0.2288 (0.2283-0.2292) & 0.1767 (0.1765-0.1770) & 0.1791 (0.1788-0.1749) & 0.1913 (0.1910-0.1915)\\
\hspace{2em}White & 0.2013 (0.2010-0.2017) & 0.2244 (0.2238-0.2249) & 0.1950 (0.1946-0.1953) & 0.1818 (0.1815-0.1821) & 0.2187 (0.2184-0.2191)\\
\hspace{2em}Black & 0.1596 (0.1582-0.1588) & 0.1906 (0.1902-0.1911) & 0.1528 (0.1525-0.1531) & 0.1481 (0.1478-0.1483) & 0.1873 (0.1870-0.1876)\\
\hspace{2em}Other & 0.2052 (0.2049-0.2056) & 0.2101 (0.2095-0.2107) & 0.1943 (0.1939-0.1947) & 0.1712 (0.1708-0.1715) & 0.2354 (0.2350-0.2358)\\
\hspace{2em}\dbs & 0.0467 (0.0464-0.0469) & 0.0337 (0.0333-0.0341) & 0.0421 (0.0419-0.0424) & 0.0231 (0.0228-0.0233) & 0.0481 (0.0479-0.0484)\\
\midrule
Age &  &  &  &  & \\
\hspace{2em}Overall & 0.1980 (0.1968-0.1992) & 0.1972 (0.1969-0.1976) & 0.1807 (0.1804-0.1810) & 0.1829 (0.1826-0.1832) & 0.1851 (0.1849-0.1854)\\
\hspace{2em}$<$ 75 yrs & 0.1766 (0.1763-0.1769) & 0.1887 (0.1883-0.1890) & 0.1726 (0.1723-0.1729) & 0.1809 (0.1806-0.1812) & 0.1784 (0.1782-0.1787)\\
\hspace{2em}$\geq$ 75 yrs & 0.2012 (0.2009-0.2016) & 0.2082 (0.2077-0.2086) & 0.1639 (0.1635-0.1642) & 0.2089 (0.2085-0.2093) & 0.1665 (0.1662-0.1668)\\
\hspace{2em}\dbs & 0.0296 (0.0294-0.0299) & 0.0195 (0.0192-0.0198) & 0.0087 (0.0084-0.0090) & 0.0280 (0.0278-0.0283) & 0.0119 (0.0117-0.0122)\\
\bottomrule
\end{tabular}
}
\end{table}

\newpage

\begin{table}[!hbpt]
\caption{The TPR and \dtpr of baseline and proposed model for thorax abnormalities diagnosis in the NIH-CXR dataset. Balance DenseNet-121 – DenseNet-121 with balanced empirical risk minimization.\cite{Vapnik1991-ro} ADV - Adversarial.\cite{Wadsworth2018-ob} SCL - supervised contrastive learning.\cite{Khosla2020-kq}}
\label{tab:tpr nih}
\resizebox{\linewidth}{!}{
\begin{tabular}{lccccc}
\toprule
& DenseNet-121 & Balance DenseNet-121 & ADV & SCL & Proposed\\
\midrule
Sex &  &  &  &  & \\
\hspace{2em}Overall & 0.4608 (0.4601-0.4615) & 0.5289 (0.5282-0.5296) & 0.4814 (0.4807-0.4820) & 0.5294 (0.5287-0.5300) & 0.4996 (0.4989-0.5003)\\
\hspace{2em}Male & 0.4645 (0.4637-0.4655) & 0.5261 (0.5252-0.5271) &   0.4824 (0.4815-0.4832) & 0.5320 (0.5311-0.5330) & 0.5115 (0.5107-0.5107)\\
\hspace{2em}Female & 0.4557 (0.4547-0.4568) & 0.5328 (0.5318-0.5338) & 0.4800 (0.4790-0.4810) & 0.5257 (0.5246-0.5267) & 0.4832 (0.4821-0.4843)\\
\hspace{2em}\dtpr & 0.0111 (0.0101-0.0122) & 0.0095 (0.0085-0.0105) & 0.0080 (0.0072-0.0088) & 0.0093 (0.0084-0.0103) & 0.0284 (0.0269-0.0298)\\
\midrule
Age &  &  &  &  & \\
\hspace{2em}Overall & 0.4608 (0.4601-0.4615) & 0.4986 (0.4978–0.4991) & 0.4799 (0.4792-0.4805) & 0.5033 (0.5026-0.5040) & 0.4837 (0.4830-0.4844)\\
\hspace{2em}$<$ 60 yrs & 0.4821 (0.4813-0.4829) & 0.5142 (0.5135-0.5150) & 0.5039 (0.5031-0.5046) & 0.5255 (0.5248-0.5263) & 0.5106 (0.5099-0.5113)\\
\hspace{2em}$\geq$ 60 yrs & 0.3835 (0.3820-0.3850) & 0.4412 (0.4397-0.4427) & 0.3927 (0.3913-0.3942) & 0.4228 (0.4212-0.4243) & 0.3860 (0.3845-0.3875)\\
\hspace{2em}\dtpr & 0.0986 (0.0969-0.1002) & 0.0730 (0.0712-0.0748) & 0.1111 (0.1095-0.1127) & 0.1027 (0.1009-0.1044) & 0.1246 (0.1230-0.1263)\\
\bottomrule
\end{tabular}
}
\end{table}

\newpage

\begin{table}[!hbpt]
\caption{The FPR and \dfpr of baseline and proposed model for thorax abnormalities diagnosis in the NIH-CXR dataset. Balance DenseNet-121 – DenseNet-121 with balanced empirical risk minimization.\cite{Vapnik1991-ro} ADV - Adversarial.\cite{Wadsworth2018-ob} SCL - supervised contrastive learning.\cite{Khosla2020-kq}}
\label{tab:fpr nih}
\resizebox{\linewidth}{!}{
\begin{tabular}{lccccc}
\toprule
& DenseNet-121 & Balance DenseNet-121 & ADV & SCL & Proposed\\
\midrule
Sex &  &  &  &  & \\
\hspace{2em}Overall & 0.1131 (0.1128-0.1134) & 0.1695 (0.1691-0.1698) & 0.1538 (0.1534-0.1542) & 0.1708 (0.1704-0.1711) & 0.1437 (0.1433-0.1440)\\
\hspace{2em}Male & 0.1093 (0.1089-0.1097) & 0.1602 (0.1597-0.1607) &   0.1421 (0.1416-0.1426) & 0.1632 (0.1627-0.1637) & 0.1466 (0.1461-0.1472)\\
\hspace{2em}Female & 0.1185 (0.1179-0.1190) & 0.1824 (0.1817-0.1830) & 0.1710 (0.1695-0.1707) & 0.1814 (0.1807-0.1821) & 0.1395 (0.1389-0.1401)\\
\hspace{2em}\dfpr & 0.0092 (0.0086-0.0099) & 0.0221 (0.0212-0.0230) & 0.0279 (0.0271-0.0288) & 0.0182 (0.0173-0.0191) & 0.0078 (0.0071-0.0085)\\
\midrule
Age &  &  &  &  & \\
\hspace{2em}Overall & 0.1131 (0.1128-0.1134) & 0.1487 (0.1483–0.1490) & 0.1690 (0.1686-0.1694) & 0.1464 (0.1460-0.1467) & 0.1424 (0.1421-0.1428)\\
\hspace{2em}$<$ 60 yrs & 0.1132 (0.1129-0.1136) & 0.1460 (0.1445-0.1464) & 0.1699 (0.1694-0.1703) & 0.1504 (0.1500-0.1508) & 0.1478 (0.1474-0.1482)\\
\hspace{2em}$\geq$ 60 yrs & 0.1127 (0.1120-0.1134) & 0.1572 (0.1563-0.1580) & 0.1663 (0.1654-0.1672) & 0.1338 (0.1330-0.1345) & 0.1257 (0.1249-0.1265)\\
\hspace{2em}\dfpr & 0.0049 (0.0046-0.0054) & 0.0115 (0.0106-0.0124) & 0.0065 (0.0058-0.0072) & 0.0166 (0.0157-0.0175) & 0.0221 (0.0212-0.0230)\\
\bottomrule
\end{tabular}
}
\end{table}

\newpage

\begin{table}[!hbpt]
\caption{The BS and \dbs of baseline and proposed model for thorax abnormalities diagnosis in the NIH-CXR dataset. Balance DenseNet-121 – DenseNet-121 with balanced empirical risk minimization.\cite{Vapnik1991-ro} ADV - Adversarial.\cite{Wadsworth2018-ob} SCL - supervised contrastive learning.\cite{Khosla2020-kq}}
\label{tab:bs nih}
\resizebox{\linewidth}{!}{
\begin{tabular}{lccccc}
\toprule
& DenseNet-121 & Balance DenseNet-121 & ADV & SCL & Proposed\\
\midrule
Sex &  &  &  &  & \\
\hspace{2em}Overall & 0.1920 (0.1917-0.1921) & 0.1956 (0.1954-0.1958) & 0.2005 (0.2003-0.2006) & 0.1962 (0.1960-0.1963) & 0.1930 (0.1928-0.1932)\\
\hspace{2em}Male & 0.1671 (0.1669-0.1673) & 0.1784 (0.1782-0.1786) &   0.1820 (0.1818-0.1822) & 0.1819 (0.1817-0.1821) & 0.1718 (0.1716-0.1720)\\
\hspace{2em}Female & 0.1546 (0.1544-0.1548) & 0.1672 (0.1670-0.1674) & 0.1715 (0.1714-0.1717) & 0.1737 (0.1735-0.1739) & 0.1644 (0.1642-0.1646)\\
\hspace{2em}\dbs & 0.0125 (0.0123-0.0127) & 0.0111 (0.0110-0.0113) & 0.0105 (0.0103-0.0106) & 0.0082(0.0081-0.0084) & 0.0074 (0.0072-0.0075)\\
\midrule
Age &  &  &  &  & \\
\hspace{2em}Overall & 0.1919 (0.1917-0.1921) & 0.1987 (0.1985–0.1989) & 0.2041 (0.2040-0.2043) & 0.1936 (0.1934-0.1938) & 0.1959 (0.1957-0.1960)\\
\hspace{2em}$<$ 60 yrs & 0.1764 (0.1763-0.1766) & 0.1854 (0.1852-0.1856) & 0.1926 (0.1925-0.1928) & 0.1799 (0.1798-0.1801) & 0.1854 (0.1852-0.1856)\\
\hspace{2em}$\geq$ 60 yrs & 0.1411 (0.1409-0.1412) & 0.1551 (0.1549-0.1553) & 0.1695 (0.1692-0.1696) & 0.1524 (0.1522-0.1526) & 0.1537 (0.1535-0.1538)\\
\hspace{2em}\dbs & 0.0354 (0.0352-0.0356) & 0.0303 (0.0301-0.0305) & 0.0232 (0.0230-0.0233) & 0.0276 (0.0274-0.0277) & 0.0287 (0.0285-0.0289)\\
\bottomrule
\end{tabular}
}
\end{table}

\newpage

\begin{table}[!hbpt]
\caption{The AUCs of DenseNet-121 and the proposed method trained on NIH and tested on MIMIC-CXR test dataset.}
\label{tab:auc cross}
\centering
\footnotesize
\begin{tabular}{lcc}
\toprule
 & DenseNet-121 & Proposed\\
\midrule
Sex &  & \\
\hspace{2em}Overall AUC & 0.6567 (0.6549-0.6585) & 0.7264 (0.7247-0.7280)\\
\hspace{2em}Male, mAUC & 0.6040 (0.6017-0.6064) & 0.6863 (0.6839-0.6885)\\
\hspace{2em}Female, mAUC & 0.7091 (0.7068-0.7113) & 0.7664 (0.7642-0.7686)\\
\hspace{2em}\dmauc & 0.1050 (0.1021-0.1080) & 0.0802 (0.0770-0.0833)\\
\midrule
Age &  & \\
\hspace{2em}Overall AUC & 0.6567 (0.6549-0.6585) & 0.7079 (0.7062-0.7096)\\
\hspace{2em}$<$ 60 yrs, mAUC & 0.6978 (0.6952-0.7003) & 0.7537 (0.7512-0.7562)\\
\hspace{2em}$\geq$ 60 yrs, mAUC & 0.6345 (0.6324-0.6367) & 0.6832 (0.6811-0.6853)\\
\hspace{2em}\dmauc & 0.0632 (0.0601-0.0663) & 0.0706 (0.0674-0.0737)\\
\bottomrule
\end{tabular}
\end{table}

\newpage

\begin{table}[!hbpt]
\caption{Relative changes and absolute changes in AUC, marginal AUC (mAUC) and difference between in mAUC (\dmauc) between baseline and the proposed method in two datasets.}
\label{tab:rel abs}
\centering
\footnotesize
\begin{tabular}{llrr}
\toprule
Dataset & Subgroups & Relative change (\%) & Absolute change\\
\midrule
MIDRC & Sex &  & \\
 & \hspace{2em}Overall AUC & -1.03 & -0.0084\\
 & \hspace{2em}Male, mAUC & -1.60 &  -0.0130\\
 & \hspace{2em}Female, mAUC & -0.25 & -0.0025\\
 & \hspace{2em}\dmauc & -68.10 & -0.0079\\
 & Race &  & \\
 & \hspace{2em}Overall AUC & -3.04 & -0.0249\\
 & \hspace{2em}White, mAUC & -0.90 & -0.0068\\
 & \hspace{2em}Black, mAUC & -4.40 & -0.0384\\
 & \hspace{2em}Other, mAUC & -1.50 & -0.0100\\
 & \hspace{2em}\dmauc & -13.5 & -0.0284\\
 & Age &  & \\
 & \hspace{2em}Overall AUC & -1.91 & -0.0156\\
 & \hspace{2em}$<$ 75 yrs, mAUC & -1.94 & -0.0194\\
 & \hspace{2em}$\geq$ 75 yrs, mAUC & 1.71 & 0.0125\\
 & \hspace{2em}\dmauc & -31.93 & -0.0319\\
\midrule
NIH-CXR & Sex &  & \\
 & \hspace{2em}Overall AUC & -0.18 & -0.0013\\
 & \hspace{2em}Male, mAUC & 0.52 & 0.0038\\
 & \hspace{2em}Female, mAUC & -1.10 & -0.0082\\
 & \hspace{2em}\dmauc & -48.89 & -0.0037\\
 & Age &  & \\
 & \hspace{2em}Overall AUC & -1.05 & -0.0077\\
 & \hspace{2em}$<$ 60 yrs, mAUC & -1.31 & -0.0098\\
 & \hspace{2em}$\geq$ 60 yrs, mAUC & -0.03 & -0.0002\\
 & \hspace{2em}\dmauc & -16.60 & -0.0085\\
\bottomrule
\end{tabular}
\end{table}

\end{document}